\documentclass[namedreferences]{kluwer}    
\usepackage{graphicx}
\newdisplay{guess}{Conjecture}

\begin{document}                                                                                   
\begin{article}
\begin{opening}         
\title{MHD jets around galactic objects} 
\author{Fabien \surname{Casse}\thanks{Fabien.Casse@obs.ujf-grenoble.fr}}
\author{Jonathan \surname{Ferreira}}  
\runningauthor{F.Casse \& J.Ferreira}
\runningtitle{MHD jets around galactic objects}
\institute{Laboratoire d'Astrophysique de l'Observatoire de Grenoble, BP 53
  F-38041 Grenoble cedex 9, France }
\date{October 15, 2000}

\begin{abstract}
We present a self-similar, steady-state model describing both a
magnetized accretion disc and a magnetohydrodynamic jet. We focus on the
role of a hot corona in such a structure. This corona enables the disc to
launch various types of jets. By considering the energy conservation, we also
present a diagnostic of the luminosity of the magnetized disc, which could
explain some observational signatures of galactic objects.  
\end{abstract}

\end{opening}           

\section{The disc-jet connection}  

In the Universe, jets of matter are observed around almost every kind of
systems where plasma accretes onto a central object. The central object can
be a young protostar, a compact galactic object or a supermassive
black hole. The jet velocity is always found to be a few times the
escaping velocity of the central object (e.g., \opencite{livio}). The most
successful model, accounting for all known properties of observed jets,
relies on a large scale magnetic field anchored on an accretion disc
(\opencite{bland}). The field extracts the disc angular momentum, thereby
allowing accretion, and accelerates a fraction of disc material to high
speeds. Moreover, the magnetic field confines the accelerating jet matter,
leading to a cylindrical collimation of the jet.  Hereafter, such systems
are called Magnetized Accretion-Ejection Structures (MAES).

In a MAES, the bipolar magnetic field pinches {\it the disc}. Thus, the only
force able to counteract both the magnetic and gravitational forces, {\it 
whithin the disc},  is the
vertical plasma pressure gradient. As a consequence, the mass loading in
the jet depends critically on this vertical equilibrium. This equilibrium
is the connection between the accretion motion and the jet. A more detailed
explaination of this phenomenon is displayed in \inlinecite{fab3}.  
\section{Entropy generation}
In all previous studies dealing with magnetized accretion disc steadily
launching jets, the energy equation was simply disregarded. It was replaced
by the assumption of isothermal or adiabatic magnetic surfaces. In the
turbulent disc however, some physical mechanisms (convection, turbulent
thermal conductivity) could provide a non-local energy transport, more
efficent than by photons. We can mimic these effects by the following equation
\begin{subequation}[arabic]
\begin{eqnarray*}
\rho T \frac{d S}{d t} = Q \nonumber
\end{eqnarray*}
\end{subequation}
\noindent where $Q$ is the difference between local heating and cooling
rates of the plasma. This function $Q$ is prescribed according to the
energy conservation (\opencite{fab2}). In a steady-state framework, this
equation can be written as
\begin{subequation} 
\begin{eqnarray*}
 Q  = \frac{\gamma}{\gamma -1}\frac{k_B}{m_p}\rho\vec{u_p}\cdot \vec{
  \nabla}T - \vec{u_p}\cdot\vec{ \nabla} P \ .
  \label{equi}
\end{eqnarray*}
\end{subequation}
where $T$ is the temperature, $P$ the plasma pressure, $u_p$ the poloidal
velocity of matter and $\gamma$ the polytropic index of plasma. Because the
gradient of plasma characteristics are almost vertical in a thin disc, one
can see that entropy generation will have a strong impact on the vertical
mass flux when the poloidal velocity is vertical (i.e. in the corona taking
place at the disc surface). In the disc, the entropy generation will be
advected with the flow onto the central object.  

This is an important point since advection-dominated accretion flows
(ADAF) only occurs in thick or slim discs ($h/r > 0.3$). Here, most of the
entropy is generated in the disc atmosphere and taken in the jet,
allowing the disc to be thin. Using this approximate form of the energy
equation, we solve the full set of MHD equations from the disc midplane
(resistive MHD regime) to the jet asymptotics (ideal MHD regime).
\section{Global energy conservation}
The global energy conservation enables us to know that the power released
by accretion $P_{acc}$ is shared between the total jet power $P_{jet}$ (MHD
Poynting flux, kinetic energy and enthalpy) and the disc luminosity
$L_D$. We measure the amount of entropy generated with a parameter $f$
  whose value is determined by the disc turbulence ($0<f\leq 1$).  It is
  noteworthy that, according to the amount of entropy, the
\begin{figure}[H]
\tabcapfont
\centerline{%
\begin{tabular}{ccc}
\includegraphics[width=2in,angle=-90]{gra2.epsi} &
\includegraphics[width=1in,angle=0]{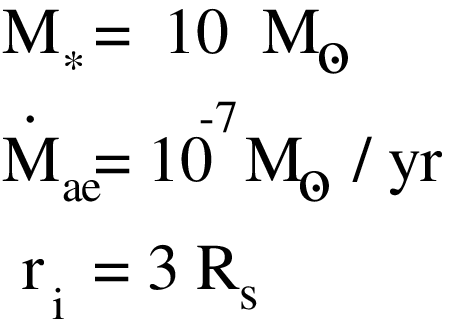} &
\includegraphics[width=2in,angle=-90]{gra1.epsi}  \\ 
a.~~ X-ray low/hard state ($f\rightarrow 1$) & &b.~~ X-ray high/soft
state ($f\rightarrow 0$)\\ 
Low disc emission \& hot corona & &Disc radiation \& no coronal emission\\
Dense hot emitting jet & &Tenuous, cold and low emitting jet 
\end{tabular}}
\caption{Poloidal cross-section of MAES. In grey are shown the
  logarithmic temperature levels, the solid lines are the logarithmic
  plasma density levels ($\log n_H = $12,12.5,13,15 in $cm^{-3}$ for left
  pannel and $\log n_H = $ 9,10,11,15 for right pannel). The denser level
  is closer to the disc. The dashed lines
  represent plasma streamlines coming from the disc to the jet. The
  poloidal magnetic field lines are not displayed here but are
  strictly parallel, in the jet, to the poloidal streamlines, according to MHD
  equations. The left 
  figure displays a structure that might be able to describe the X-ray
  low/hard state of some microquasars (see \opencite{Rob}). The right
  pannel could correspond to a X-ray high/soft state.} 
\label{1}
\end{figure}
\noindent resulting jet is magnetically-driven (``cold'' jet $f\sim0$) or
  magneto-thermally-driven (``hot'' jet $f\sim 1$). The energy conservation
  becomes (\opencite{fab2}) 
\begin{eqnarray*}
2L_D = (1 - f) P_{diss} = P_{acc} - 2P_{jet}  
\end{eqnarray*}
where $P_{diss}$ is the viscous and ohmic heating integrated all over the
disc. One can easily see that if most of heat is deposited as entropy in
the corona, a thin magnetized disc could launch jets and have a weak
luminosity. Assuming the disc to be optically thick, we can calculate its
luminosity $L_D$ and effective temperature $T_{eff}$, namely
\begin{eqnarray*}
\frac{L_D}{L_{EDD}} &\simeq&
0.36\left(\frac{1-f}{1+\Lambda}\right)\left(\frac{\dot{M}_{ae}}{10^{-7}M_{\odot}/yr}\right)\left(\frac{M_*}{10M_{\odot}}\right)\left(\frac{3R_s}{r_i}\right)\\
kT_{eff} &\simeq& 1
keV\left(\frac{1-f}{1+\Lambda}\right)^{\frac{1}{4}}\left(\frac{\dot{M}_{a}(r)}{10^{-7}M_{\odot}/yr}\right)^{\frac{1}{4}}\left(\frac{M_*}{10M_{\odot}}\right)^{-\frac{1}{2}}\left(\frac{r}{3R_s}\right)^{-\frac{3}{4}}
\end{eqnarray*}
where we have assumed the external radius $r_e$ of the MAES to be much
larger than the internal radius $r_i$ ($R_s=2GM_*/c^2$  and
$L_{EDD}$ is the Eddington luminosity). The
parameter $\Lambda$ is the ratio of the magnetic to the viscous
torque at the disc midplane and depends on the local MHD turbulence
(\opencite{fab3}).  

An important consequence arises from these estimates. Indeed a magnetized
disc driving jets could be less luminous than a standard accretion disc,
even with $f\simeq 0$ (no corona). This will be the case if the turbulence
is such that the magnetic torque is larger than the viscous one, the disc
will be darker but keeping almost the same effective temperature. 

In a stationary MAES, the characteristic timescale for jet launching is only a
few dynamical timescales, i.e. roughly one second around a stellar black
hole. So both low and high state of microquasars (e.g. \opencite{Mir},
 \opencite{Rob}) might be described by two different MAES. Indeed
the high state exhibits a strong thermal emission in X-rays (soft X-rays
emitted from the disc, $f\rightarrow 0$) and a very small radio emission
that could be accounted by a very tenuous, cold jet (or no jet at all). At the
opposite, the low state is characterized by a non-thermal emission in hard
X-rays (associated with a corona) and a ``radio'' jet. This low state might
be obtained with a MAES where most of the heat is not released as radiation
by the optically thick disc but shared between coronal
losses and thermal energy sent into the jet ($f\rightarrow 1$). The
resulting jets are hot and dense, and may be associated with a radio and
infra-red emissions. In Figure \ref{1}, we display these two MAES. 
Anyway, a disc instability allowing the MAES to shift between these two
states remains to be found. Also, to clearly validate this scenario, a
modelisation of spectrum emitted from MAES must be performed and compared
to observations. The authors would like to thanck Rob Fender for fruitful
discussions.  

\end{article}

\begin{thebibliography}{}
\bibitem[\protect\citeauthoryear{Blandford \& Payne}{1982}]{bland}
Blandford, R.D. and Payne, D.G., 1982, {\it MNRAS\/} Vol.199 pp. 883
\bibitem[\protect\citeauthoryear{Casse \& Ferreira}{2000a}]{fab3}
Casse, F. and Ferreira, J., 2000a, {\it A\&A\/} Vol.353 pp. 1115-1128
\bibitem[\protect\citeauthoryear{Casse \& Ferreira }{2000b}]{fab2}
Casse, F. and Ferreira, J., 2000b, {\it A\&A\/} Vol.361 pp. 1178-1190 
\bibitem[\protect\citeauthoryear{Fender et al.}{1999}]{Rob}
Fender et al., 1999, {\it ApJ\/}  Vol.519 pp. L165-L168 
\bibitem[\protect\citeauthoryear{Livio}{1997}]{livio}
Livio, M., 1997: ``The formation of astrophysical jets'' {\it ASP
  conferences series\/} Vol.121, Wickramasinghe T., Ferrario L.,
  Bicknel, G.V.(eds.), p.845 
\bibitem[\protect\citeauthoryear{Mirabel et al.}{1998}]{Mir}
Mirabel et al., 1998, {\it A\&A\/} Vol.330 pp. L9-L12
\end{thebibliography}
\end{document}